# Application Location Based Service (LBS) Location Search Palembang Nature-Based Android

Intan Okta Sari[1], Leon Andretti Abdillah[2], Kiky Rizky Nova Wardhani[3]
[1] Informatics Engineering, Bina Darma University
[2,3] Information Systems, Bina Darma University
Jalan Ahmad Yani No.3, Plaju, Palembang
[1] intansari.12142243@gmail.com, [2] leon.abdillah@yahoo.com

**Abstract.** With the development of information systems to make the operating system more diverse mobile devices, the emergence of the Android operating system that is open allows users to search for and acquire various information easily and quickly. Application Search Nature Places is an application that can help bring information on nearby Places Nature is all around. Can be used in the Android Operating System and Global Positioning System (GPS). To be able to use this application, users must be connected to the Internet because it requires data taken from Google Maps. The main facilities contained in this application is a feature that makes it Map and Route users in finding the intended location. With the LBS application is expected to provide information that is accurate, clear and precise to determine location points Nature Palembang, and can facilitate local and foreign tourists and the public, especially the city of Palembang.

**Kyewords:** Location Based Service (LBS), Android, Global Position System (GPS), Travel.

## 1 Introduction

The rapid development of mobile technology in line with the development of information technology (IT). It becomes an opportunity for the developers of IT. One of the most progreessive applications on IT recently is mobile technology [1]. By using the mobile device information can be obtained easily in a short time. Mobile applications alredy used for : 1) Al Quran mobile learning [2], 2) Bus ticket reservation application [1], 3) Mobile dictionary [3], 4) GIS-based residential locations [4], or 5) Geographic Information System of Urban Green Open Space [5]. Various mobile phone has many uses the Android operating system. Android is a package of software for mobile devices, including an operating system, middleware and core applications [6]. Android have been used in billion of gadgets like smatphones and tablets [3].

With the development of technology that is increasingly advancing a positive influence for Android users and one of them with the use of Global Positioning System (GPS). GPS technology enables geographical information (latitude and longitude coordinates) [7]. GPS allows the development of Location Based Service (LBS). LBS uses Google Maps API to implement mobile mapping services [8]. LBS is one service that actively act to change the position of the entity so that it can detect the location of objects and provides services according to the location of the known objects. Location based Services give current users' location to retrieve more valuable information near to their location [9].

Most people have to rely on mobile devices to obtain information, including information about nature. That's why the writer android based LBS application development is expected to provide clear information about the location points Palembang travel and able to assist local tourists, foreign and especially the people around Palembang. LBS applications built using the Eclipse software, the Java SDK and Android SDK. Excellence LBS applications are specifically structured to 6 (six) tourist sites Nature Palembang city and the starting point is automatically detected using GPS and Google maps.

The rest of this article are consists of: 1) research methodology as section 2 (two), 2) result and discussions as section 3 (three), and 3) conclude with conclussions as section 4 (four).

## 2  Research Methodology

This research was conducted at the Department of Culture and Tourism of Palembang. Methods of data collection in this study were: 1) observation, a method of data collection where researchers recorded information as they witnessed during the study. In this case I made some observations to the Department of Culture and Tourism of the city of Palembang. 2) Documentation, is a method of collecting data by finding and collecting data on issues related to research. 3) Study Library (Literature), is a method of data collection conducted by researchers with examining theories, opinions and support the main ideas to the issues discussed.

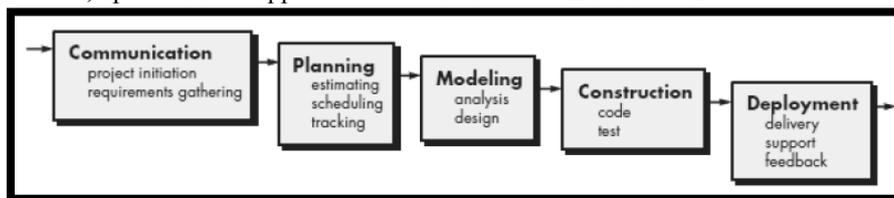

**Fig. 1.** Waterfall Model.

The development method used in this study is a Waterfall model [10]. This model is a model of systematic approach that sequentially in software development that begins with the specification needs - needs and will start from the analysis of user needs continue into the process of designing, coding, testing, and maintenance gradually, Waterfall consists of five (5) phases: 1) Communication, 2) Planning, 3) Modeling, 4) Construction, and 5) Deployment (Picture 1)

## 3 Results and Discussions

Based on the research that has been done, the design and ends with the creation of actual program, the results achieved by the authors is an Application Location Based Service (LBS) Location Search Nature Palembang Based on Android, this application consists of: Splash Screen, Main Page, Wisata Alam, About, and Exit.

### 3.1 Splash Screen and Main Page

Screen Splash pages are beginning to see applications that are shown before the main page. On page splash screen will display the icon Ampera bridge. Page splash screen (figure 2) will run for 3 (three) seconds before the main page (figure 3) of the application is displayed.

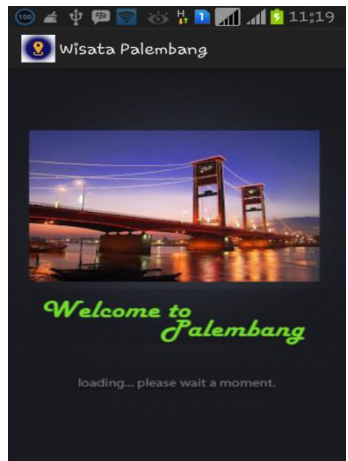  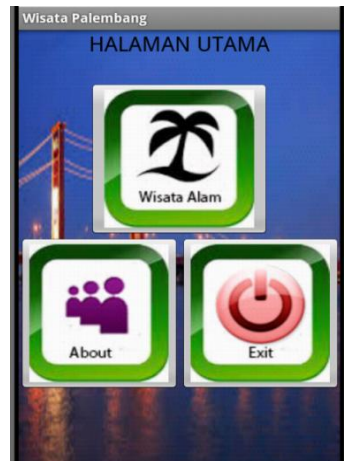

**Fig. 2.** Splash Screen.   **Fig. 3.** Main Page.

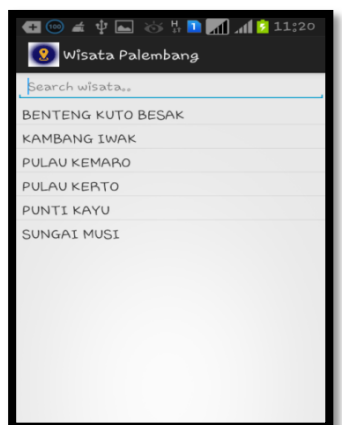

**Fig. 4.** Natural Tourism.

### 3.2 Natural Tourism

This is the main page that list whole natural tourism in Palembang (figure 4). This page/menu display the natural tourism destinations, such as : 1) Kuto Besak, 2) Kambang Iwak, 3) Kerto Island, 4) Island Kemaro, 5) Punti timber, and 6) Musi River (see figure 4).

### 3.3 Description and Route Maps Nature

Description page and route nature (figure 5) is the detail page of the menu pages nature, this page will be displayed when the user presses one listview displayed by the page menu nature.

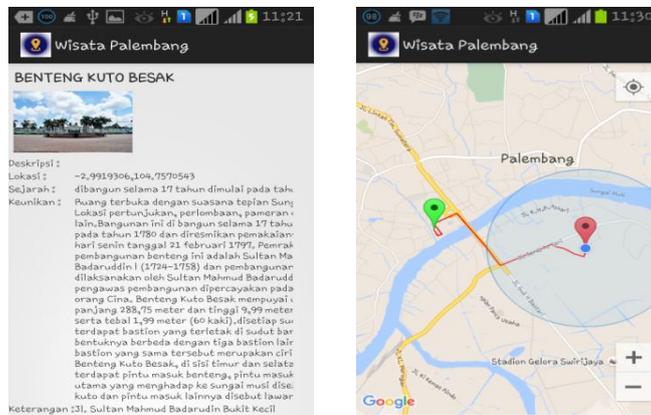

**Fig. 5.** Page Description and Route Maps of Natural Tourism.

Figure 5 is the example one of natural tourism of Benteng Kuto Besak. This page consists of the description of the particular natural tourism. This page then shows the position of current user (red pointer in the circle), and a line that linked inked into destination spot (green pointer).

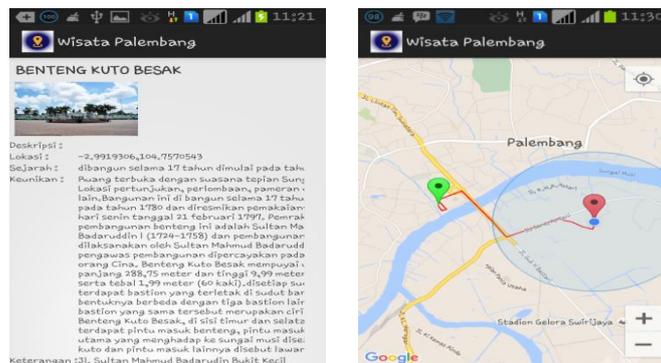

**Fig. 6.** Page Description and Route Maps of Natural Tourism.

### 3.4 Administrator Page

Administrator system software is software that is in the web server. System administrators will be used by administrators to process the input, edit and delete menu that will be displayed in LBS applications in the android operating system. The login page is the initial view of the system administrator when accessed. Before heading to the main page of the system administrator, the admin must first input the appropriate username and password. If the username and password as the system administrator will display the main page. Admin main page is a page that is used to add a menu to be displayed on android based mobile LBS application. Natural attractions page is a page that will display data from menu natural attractions that have been added by the admin. Natural attractions input page is a page that will display the data nature LBS applications to be written by admin.

## 4 Conclussions and Recommendations

Based on the research that has been done in building applications Location Based Service (LBS) search the natural attractions of the city of Palembang-based android, it can be concluded that:
1) This research resulted in an application Location Based Service (LBS) search the natural attractions of the city of Palembang-based android that can be run on the Android operating system with a minimum version 2.2 (Froyo), up to the latest version 5.0 (Lollipop).
2) LBS applications built using the Java programming language to build software on android mobile devices.
3) The software can be accessed using the Internet connection or online because the software uses techniques JSON (JavaScript Object Notation) as a data exchange format that connects the database which resides in webserver with mobile android software and integrated with Google Maps.

## References

1. Fitriani, et al., "Android-based bus ticket reservation application," in 4th International Conference on Information Technology and Engineering Application 2015 (ICIBA2015), Palembang, 2015.
2. M. Sobri and L. A. Abdillah, "Aplikasi belajar membaca iqro' berbasis mobile," in Seminar Nasional Teknologi Informasi & Multimedia (Semnasteknomedia), STMIK AMIKOM Yogyakarta, 2013.
3. Murdianto, et al., "Dictionary of prabumulih language-based android," in 4th International Conference on Information Technology and Engineering Application 2015 (ICIBA2015), Palembang, 2015.
4. L. N. Sari, et al., "Geographic information systems of android-based residential locations," in 4th International Conference on Information Technology and Engineering Application 2015 (ICIBA2015), Bina Darma University, Palembang, 2015.
5. Andika, *et al.*, "Sistem Informasi Geografis Ruang Terbuka Hijau Kawasan Perkotaan (RTHKP) Palembang," presented at the Student Colloquium Sistem Informasi & Teknik Informatika (SC-SITI) 2015, Palembang, 2015.
6. X. Shu, et al., "Research on mobile location service design based on Android," in Wireless Communications, Networking and Mobile Computing, 2009. WiCom'09. 5th International Conference on, 2009, pp. 1-4.
7. H. Q. Vu, et al., "Exploring the travel behaviors of inbound tourists to Hong Kong using geotagged photos," Tourism Management, vol. 46, pp. 222-232, 2015.
8. O. A. Ibrahim and K. J. Mohsen, "Design and Implementation an Online Location Based Services Using Google Maps for Android Mobile," International Journal of Computer Networks and Communications Security (CNCS), vol. 2, pp. 113-118, 2014.
9. M. Singhal and A. Shukla, "Implementation of location based services in Android using GPS and Web services," IJCSI International Journal of Computer Science Issues, vol. 9, pp. 237-242, 2012.
10. R. S. Pressman, Software Engineering: A Practitioner's Approach, 7th ed. New York, US: McGraw-Hill, 2010.